# AUTOMATIC DETECTION OF ACCESS CONTROL VULNERABILITIES VIA API SPECIFICATION PROCESSING

Alexander Barabanov[1], Denis Dergunov[2], Denis Makrushin[3], Aleksey Teplov[4]

**Abstract**

**Objective**. Insecure Direct Object Reference (IDOR) or Broken Object Level Authorization (BOLA) are one of the critical type of access control vulnerabilities for modern applications. As a result, an attacker can bypass authorization checks leading to information leakage, account takeover. Our main research goal was to help an application security architect to optimize security design and testing process by giving an algorithm and tool that allows to automatically analyze system API specifications and generate list of possible vulnerabilities and attack vector ready to be used as security non-functional requirements.
**Method**. We conducted a multivocal review of research and conference papers, bug bounty program reports and other grey sources of literature to outline patterns of attacks against IDOR vulnerability. These attacks are collected in groups proceeding with further analysis common attributes between these groups and what features compose the group. Endpoint properties and attack techniques comprise a group of attacks. Mapping between group features and existing OpenAPI specifications is performed to implement a tool for automatic discovery of potentially vulnerable endpoints.
**Results and practical relevance**. In this work, we provide systematization of IDOR/BOLA attack techniques based on literature review, real cases analysis and derive IDOR/BOLA attack groups. We proposed an approach to describe IDOR/BOLA attacks based on OpenAPI specifications properties. We develop an algorithm of potential IDOR/BOLA vulnerabilities detection based on OpenAPI specification processing. We implemented our novel algorithm using Python and evaluated it. The results show that algorithm is resilient and can be used in practice to detect potential IDOR/BOLA vulnerabilities.

# 1 Introduction

The microservice architecture is being increasingly used to design applications system that could process confidential data and provide critical functionality [16], [17], [18], [19]. One of the most critical vulnerability related to the distributed nature of microservice-based systems is "Insecure Direct Object Reference" (IDOR) or "Broken object level authorization" (BOLA) [15]. Therefore, application security architects have to apply different methods and tools during secure software development lifecycle or secure development lifecycle (SDL) to mitigate IDOR/BOLA risks and, on the other hand, security researchers should investigate that research area and create novel methods, tools and techniques to detect such type of vulnerability.

It should be noted that nowadays development teams usually uses "API first"[5] approach – it involves developing APIs that are consistent and reusable, which can be accomplished by using an API description language to establish a contract for how the API is supposed to behave. Our motivating question was: is it possible to use API specification (e.g. written in OpenAPI format) to automatically (or semi-automatically) derive security non-functional requirements on the SDL design phase (threat modeling process) to address potential access control vulnerabilities? Therefore, our main research goal was to help an application security architect to optimize workflow by giving an algorithm and tool that allows us to automatically analyze API specifications and generate list of possible vulnerabilities and attack vector ready to be used as security non-functional requirements in SDL design phase.

---

[1] Alexander Barabanov, Ph.D, CISSP, CSSLP, Associate Professor, Department of Information Security, Bauman Moscow State Technical University, Moscow, Russia. E-mail: barabanov.iu8@gmail.com
[2] Denis Dergunov, Research Engineer, Advanced Software Technology Laboratory, Huawei, Moscow, Russia. E-mail: dan.dergunow@gmail.com
[3] Denis Makrushin, OSCP, Chong-Ming Technology Center, Director, Huawei, Moscow, Russia. E-mail:denis@makrushin.com
[4] Aleksey Teplov, Ph.D, Big Data Team Technical Leader, Advanced Software Technology Laboratory, Huawei, Moscow, Russia. E-mail: aleksey-teplov@huawei.com

[5] Joshua S. Ponelat and Lukas L. Rosenstock. Designing APIs with Swagger and OpenAPI, 2021. https://www.manning.com/books/designing-apis-with-swagger-and-openapi

E.g., given the following endpoint description in OpenAPI format (Figure 1) we can automatically derive that endpoint "showPetById" can potential be attacked using "petId" parameter manipulation in URL. Then we can propose development team with concrete non-functional requirement to address that threat: "Ensure that microservice enforce access control while processing HTTP GET request on /pets/{petId} endpoint by checking that requested petId belongs to user ID from request".

```
openapi: "3.0.0"
…
paths:
  /pets/{petId}:
    get:
      summary: Info for a specific pet
      operationId: showPetById
      tags:
        - pets
      parameters:
        - name: petId
          in: path
          required: true
          description: The id of the pet to retrieve
          schema:
            type: string
```

*Figure 1* OpenAPI specification example

To reach our research goal, we provided systematic analysis and review of research papers as well as presentations/bug reports/bug bounty tips about IDOR/BOLA-related attacks. We synthesized groups of the attacks based on their commonalities. Then we made an analysis of reviewed attacks to outline attack techniques used, properties of a target that led to possibility of vulnerability to exist, and relations between target properties and possible techniques to apply in an attack vector (checker rules). Finally, we developed an algorithm of OpenAPI specification processing that complements existing information about endpoints with IDOR/BOLA related properties. We developed an algorithm that takes complemented information of an endpoint and decides whether it is potentially vulnerable to IDOR/BOLA-based attacks and gives specific descriptions of what attack techniques and requests might be successful to exploit a vulnerability if it exists. We developed an API specification scanning tool based on OpenAPI version 3 specification standard that implements proposed novel algorithms and tested our algorithms on real case OpenAPI specifications of a service.

In summary, this paper makes the following contributions:

- We provide systematization of IDOR/BOLA attack techniques based on literature review and real cases analysis and derive IDOR/BOLA attack groups (Section 2);
- We proposed an approach to describe IDOR/BOLA attacks based on OpenAPI specifications properties (Section 3);
- We develop an algorithm of potential IDOR/BOLA vulnerabilities detection based on OpenAPI specification processing (Section 4);
- We implemented our novel algorithm using Python and evaluated it. The results show that our algorithm is sound and practical in potential IDOR/BOLA vulnerabilities detection (Section 5).

# 2 Systematization of IDOR/BOLA attack techniques based on literature review

In the first step, we gathered information on existing IDOR/BOLA attack vectors by analyzing academic and gray literature, existing testing tools, bug bounty programs and CVE reports. Grey literature includes presentations at the major security conferences, publications made in blogs, videos and company websites. Grey literature might be a valuable source when providing an example of an attack request, or report that is not disclosed on the bug bounty programs due to any reason.

To find research articles in the academic literature, we used Google Scholar and queried major digital databases: ScienceDirect, ACM Digital Library, IEEE Xplore, Springer Link. To find information in bug bounty programs or CVE reports we used generic web search engines with advanced operators for building structured queries. Because CVE is usually linked to CWE, while reviewing CVE report we analyzed CVE reports with CWE-639. We also used general web search engines with structured query strings and advanced operators to delimit output for certain CWEs and keywords like IDOR, BOLA, etc.

It should be mentioned that there is no consensus on terminology used to describe the type of vulnerability we are interested in. In addition, one can bear in mind a broader or narrower set of vulnerabilities under each term based on their preferences. Therefore, we used multiple keywords in our structured query strings: "IDOR/Insecure direct object reference", "BOLA/Broken object level authorization", "Improper Access Control" and "Broken Access Control".

We provided literature review in August –September 2021. To delimit the results, the following criteria were applied for the search results:

- publication date is since 2016;
- article/report contains request example or detailed description at the best or brief description to check IDOR/BOLA vulnerability.
- described attack is targeted over vulnerable HTTP parameter or parameter in JSON schema passed in HTTP request body.

As a result of the literature review we got the following result.

1) We were unable to find a huge amount of academic literature dedicated to IDOR/BOLA vulnerabilities containing (two publications were found).
2) Many CVE reports do not have description but only links to external sites where information in any form is presented. Not all vulnerabilities have a proper CWE assignment: most of IDOR vulnerabilities are CWE-639 or CWE-22 but some of these have CWE-20, CWE-869 and so on. Probably we can find more CVE using reverse search techniques on CWE-639, or CWE-20, CWE-285. Not all of these CVEs are about IDORs in HTTP.
3) Bug bounty program "HackerOne": there are considerable amounts of public reports with fully and partially disclosed details. Significant part of reports gives full understanding of attack vector techniques and properties.
4) Bug bounty program "BugCrowd": there are no disclosures of IDORs except for internal CTF challenges summaries.
5) Bug bounty program "YesWeHack": there are no disclosures at all, only streaming log of CWE-tags of found vulnerabilities is available. Lots of CWE-639/IDORs but cannot view disclosures.
6) We analyzed personal and company's blogs and articles about vulnerabilities disclosures with attack or exploited detailed description. Some disclosures do not have assigned CVE or report on bug bounty platform (or report is published without proper tags we searched for) but still consider real enterprise sites or applications (Facebook, Twitter, Instagram, etc.)

7) We identified and analyzed several security tools focused to detect IDOR/BOLA vulnerabilities: RESTler (Microsoft)[6], Autorize (Burp extension)[7], fuzz-lightyear (Yelp)[8].

Then we analyzed gathered information in order to systematize existing IDOR/BOLA attack techniques. On the table (Table 1) below we present systematization of IDOR/BOLA attack techniques based on provided literature review (where "CVE" is CVE reports, "BB"- bug bounty reports, "AL" – academic literature, "SW" – software tools, "GL" – gray literature).

*Table 1*

| IDOR/BOLA attack group | Source | | | | |
|---|---|---|---|---|---|
| | CVE | BB | AL | SW | GL |
| Single numerical/sequential ID in URL enumeration without a priori knowledge (dumb) | + | | | | + |
| Single numerical ID in body enumeration without a priori knowledge (dumb) | + | + | | | + |
| Single numerical ID Body (JSON) enumeration without a priori knowledge (dumb) | | + | [8] | | + |
| Single numerical ID resource path enumeration without a priori knowledge (dumb) | | + | | | + |
| Single numerical/sequential ID in URL and Body's JSON object parameter pollution (without a priori knowledge) | + | + | | | |
| Single parameter account information in Resource Path and Body (JSON) parameter pollution | | + | | | |
| Single numerical ID in header enumeration with a priori knowledge (gray box, sequence of requests) | + | | | | |
| Single parameter account information in body (JSON) enumeration with a priori knowledge | | + | | | |
| Single parameter UUID in body (JSON) enumeration with a priori knowledge | | + | | | |
| Single numerical ID in body (JSON) enumeration with a priori knowledge | | | | | + |
| Single numerical ID in resource path enumeration with a priori knowledge | + | | [2] | | |
| Multiple parameters: numerical ID, numerical ID in body (JSON) enumeration with a priori knowledge | | | | | + |
| Multiple parameters: String (from enum), numerical ID in URL enumeration without a priori knowledge | + | | | | |
| Authorization cookie swap (Dumb/Chained or sequenced) | + | | [2] | + | |

# 3 Approach to describe IDOR/BOLA attacks based on OpenAPI specifications properties

Because our main idea is to parse OpenAPI specification and derive potential IDOR/BOLA vulnerabilities we provided analysis of possible OpenAPI specifications properties used to describe endpoints in order to understand what properties and its values could be an evidence of potential IDOR/BOLA. We provided OpenAPI parameters analysis for the following cases: endpoint properties, endpoint and method properties, parameter properties. Then we derive attributes and its possible values that can be evidence of potential IDOR/BOLA vulnerability (Table 2).

---

[6] https://github.com/microsoft/restler-fuzzer
[7] https://github.com/portswigger/autorize
[8] https://github.com/Yelp/fuzz-lightyear

| Attribute name | Possible value | Value description |
|---|---|---|
| Endpoint properties | | |
| Defined HTTP verbs | All | Every HTTP verb is defined |
| | Multiple | Endpoint does not specify all HTTP verbs so request without defined verb must return 405 or 501 error. |
| | Single | Only one HTTP verb is defined |
| Endpoint and method properties | | |
| Method/Operation parameters | Empty | Endpoint does not specify parameters and its operations too |
| | Endpoint-level only | Only endpoint-level parameters used in operation, so set of parameters used in different operation may be the same |
| | Operation has parameter besides endpoint-level (non-empty) | Operation defines its own parameters to use in conjunction with endpoint level, so it is meaningful to send request with one verb using parameters list of another verb |
| Number of identifiers/parameters targeted/affected | Zero | HTTP request does not contain parameters related to object identification. |
| | One/Single | HTTP request contains parameters related to object identification. One parameter is affected and manipulated to send more requests with the parameter's value changed to check IDOR vulnerability. |
| | Multiple | HTTP request contains more than one parameter related to object identification. Several parameters are affected and manipulated at one time during a sequence of requests to check IDOR vulnerability. |
| Authorization required | Yes/True | Authorization is required, so authorization checks must be performed and potential IDOR might be in place |
| | No/False | Authorization is not used so no need to find IDORs. |
| Parameter properties | | |
| Parameter location | Resource path in URI | It is common to have request's URI in the form of origin and resource path and parameter set. Resource path components are strings like: "/resourceName", "/resourceID" and "/action". These strings would constitute a group and the resource path is a sequence of such groups. Attack vectors can be targeted over these groups and tamper with resourceID, manipulating with resourceName (case conversion), switching action or insertion of ../../ as for Local file Inclusion vulnerability. All these actions are to bypass authorization. |
| | URL parameter | Object identifier is a parameter in a query string. For example, GET requests should not have a body and put request's parameters in URI. |
| | Body | Object identifier is a parameter placed in a request's body. |
| | JSON (in body) | Parameters stored in a JSON object written in the request's body |
| | Request Header | There are found reports about IDOR vulnerability with tampered parameter in Header section |
| ID Type | Numerical sequential Identifier | Object's identifier is a number and possibly decorated with some constant substrings. Only the number part of the identifier is significant and can be modified to identify another object. Small increase or decrease of the number would identify objects created |
| | | in nearly the same time. Enumeration is the easiest way to request other existing objects. |
| | UUID/GUID | Object identifier is some variant of UUID. Usage of UUID helps to protect endpoints from enumeration attack because possibility of "guessing" object UUID in orders of magnitude lower compared to ordinary numerical identifier. |
| | Account/Personal information | Object identification uses non-obfuscated identifiers like e-mail address, phone number etc. Such information can be stored in public or received from other endpoints leading to some form of enumeration. |
| | Array | Parameter is an array of identifiers of resources. Identifiers of resources a user not permitted to access can be appended to the list of expected identifiers to |
| | String | Identifier is a string literal that can't be recognized as UUID/GUID and uses custom pattern or structure |
| | Other | When identifier's type can't be recognized with proposed heuristic rules |

On the next step we provide analysis of identified attributes and its possible values and identified IDOR/BOLA attack group in order to define the set of IDOR/BOLA attack vector techniques described using attributes (see Column "Condition" in Table 3) derived from OpenAPI specification (Table 3). We define 4 groups of IDOR/BOLA attack vectors and 10 IDOR/BOLA attack techniques overall.

*Table 3*

| Group | Technique | Technique description | Condition |
|---|---|---|---|
| Enumeration | Enumeration without a priori knowledge | Identifier is tampered for enumeration based on automatically or semi-automatically determined pattern. In the simplest form, identifier is sequential and enumeration leads to targeting existing object with identifier being unknown at the start | Parameter's type is Numerical |
| | | | Operation uses authorization AND operation parameters is not Empty AND number of identifiers/parameters targeted/affected is NOT zero AND |
| | Enumeration with a priori knowledge | Targeted identifier structured in a way that it's hard to automatically enumerate it but still needed to check with a set of known identifiers of non-owned objects. In combination with information disclosure vulnerability, impact of BOLA increases because an attacker would exploit vulnerability without brute-forcing techniques | Parameter's type is UUID/GUID, JSON object, encoded object |
| | | | Operation uses authorization AND operation parameters is not Empty AND Number of identifiers/parameters targeted/affected is NOT zero AND parameter's type is "complex" |
| | | | Parameter's type is string |

| | Add/Change file extension | A variation of enumeration when enumerated identifier is appended with an extension or changed to another extension | Operation uses authorization AND operation parameters is not Empty AND number of identifiers/parameters targeted/affected is NOT zero |
|---|---|---|---|
| | Wildcard (*,%) replacement/appending | A variation of enumeration when enumerated identifier is decorated with a wildcard or a special character | Parameter's type is string |
| | | | Operation uses authorization AND operation parameters is not Empty AND number of identifiers/parameters targeted/affected is NOT zero |
| | ID encoding/decoding | A variation of enumeration when not only an encoded identifier is enumerated but a decoded identifier is substituted too | Operation uses authorization AND operation parameters is not Empty AND number of identifiers/parameters targeted/affected is NOT zero |
| | (JSON) List appending | Parameter's type is array/list with one or few values and identifiers of non-owned objects are appended to that list to exploit improper access control | Parameter's type is array |
| | | | Operation uses authorization AND operation parameters is not Empty AND number of identifiers/parameters targeted/affected is NOT zero AND |
| Authorization token manipulation | Authorization token manipulation | Request is repeated with authorization cookies of another user to check whether authorization is incorrect | Operation uses authorization (security field is not empty or equal to [ ]) (overriding top-level declaratory) |
| Parameter pollution | Parameter pollution | Information in one request is processed and sent into different processing units of server. Tampering with one of parameter's value is a way to check that authorization is consistent and there's no case that value from one | Two parameters with same name but different locations:<br>• Resource path<br>• URL<br>• Body<br>• JSON in Body |

| | | location is used for authorization and value from another is used to access an object | Operation uses authorization AND Number of identifiers/parameters targeted/affected is Multiple |
|---|---|---|---|
| Endpoint verb tampering | Adding parameters used in other HTTP Methods | Authorization may be performed for a concrete verb and its parameters but service logic ignores requests verb | Endpoint has more than two methods, parameter set is not the same |
| | | | Defined HTTP endpoints property's value IS NOT single AND Operations' parameters list are not same or empty AND at least one of operations uses authorization |
| | Change HTTP Method (Verb tampering) | Request's verb is changed to another verb that is not specified in the endpoint's description. Incorrect behavior is when authorization checks are performed over described verbs and verb transformation is performed after authorization check (PUT->POST) | Defined HTTP endpoints property's value is not 'all' AND Number of identifiers/parameters targeted/affected is not zero |

# 4 Algorithm of potential IDOR/BOLA vulnerabilities detection based on OpenAPI specification processing

In that section, we propose our novel algorithm to parse OpenAPI specification in order to identify potential IDOR/BOLA vulnerabilities. The algorithm comprises two main stages (Figure 2).

1) Stage "IDOR/BOLA properties analyzer". Take a valid OpenAPI specification and define values of attributes related to potential IDOR/BOLA vulnerabilities described in Section 3. Such values are written as fields in input OpenAPI objects associated with endpoints, methods and parameters. We call that output augmented specification as annotated OpenAPI specification.
2) Stage "Attack analyzer". Take annotated OpenAPI specification and determine what attack vector techniques are applicable. This is performed based on value of attributes in annotated specification and usage condition of attack vector technique described in Section 3. If it is found that some attack vector technique's condition is satisfied then we consider a potential IDOR/BOLA vulnerability to be detected and specify a combination of endpoints, operations and parameters that are potentially vulnerable and have to be targeted with corresponding attack vectors.

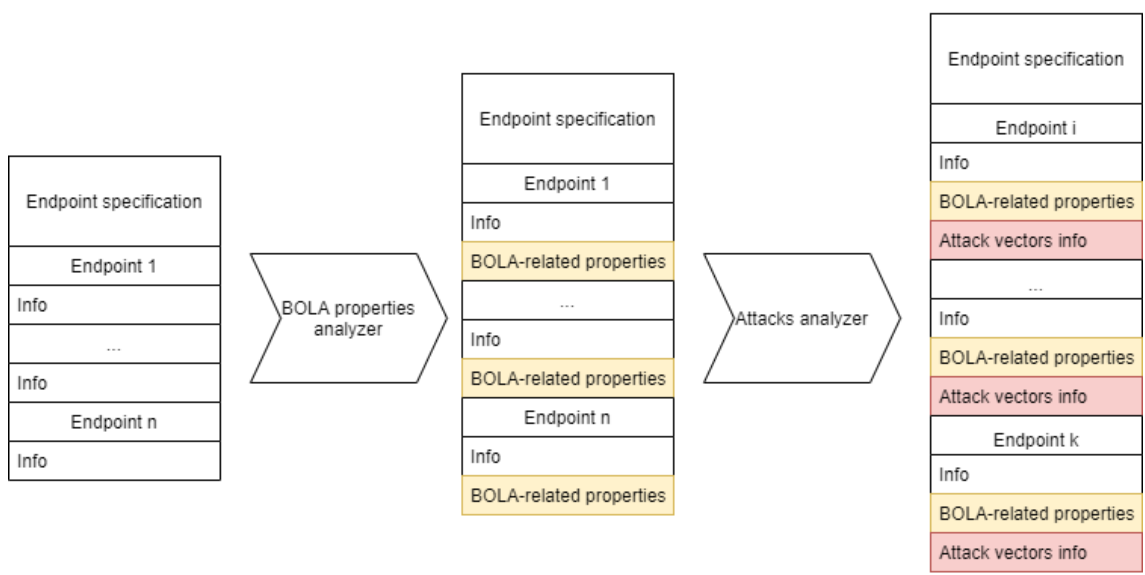

*Figure 2. The algorithm to parse OpenAPI specification in order to identify potential IDOR/BOLA vulnerabilities*

On the stage "IDOR/BOLA properties analyzer" (Figure 3) our algorithm takes OpenAPI specification as an input and annotates it in the following way[9].

1) Parse OpenAPI specification and check security authorization scheme.
2) If the OpenAPI object's paths field is empty, then finish the process, else proceed to step 3.
3) Determine order of individual endpoints processing based on corresponding path patterns. Proceed with analysis of the first endpoint's "path item" object.
4) If "path item" object contains "parameters field, then annotate each "parameter" object in the list with parameter level properties from Table 2
5) If "path item" object contains "operation" object, then proceed with analysis of each them as follows:
    a. If "operation" object contains parameters field, then annotate each "parameter" object in the list with parameter level properties from Table 2;
    b. Annotate "operation" object with method level properties from Table 2.
6) Annotate "path item" object with endpoint level properties from Table 2.
7) Proceed with processing of the next "path item" object as described in steps 4-6. If all endpoints are processed, save annotated specification.

---

[9] In OpenAPI terminology: each individual endpoint is described by a "path item" object; the endpoint's method (for instance, GET or POST) is described by an "operation" object; and a request's parameter is described by a "parameter" object.

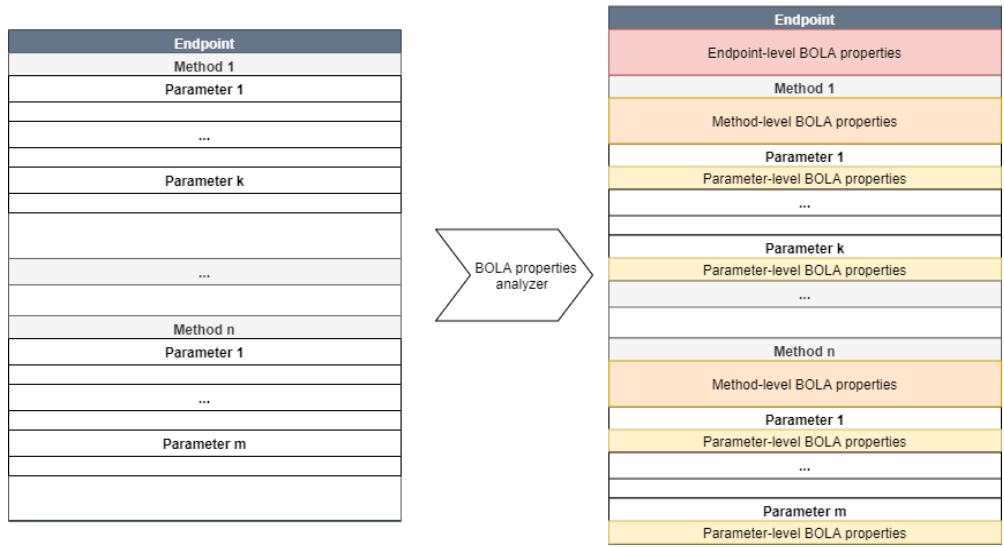

*Figure 3.* Structure of endpoint's path item object in OpenAPI specification before and after annotating it with IDOR/BOLA-related properties

On the stage "Attack analyzer", the algorithm takes an annotated OpenAPI specification with defined values of IDOR/BOLA-related properties and searches for a combination of endpoint specification suitable to attack using prepared attack vector patterns. Each pattern is specified by an endpoint, set of operations and their parameters resulting in potential vulnerability detection with attack vector proposal. The process can be described as follows (Figure 4).

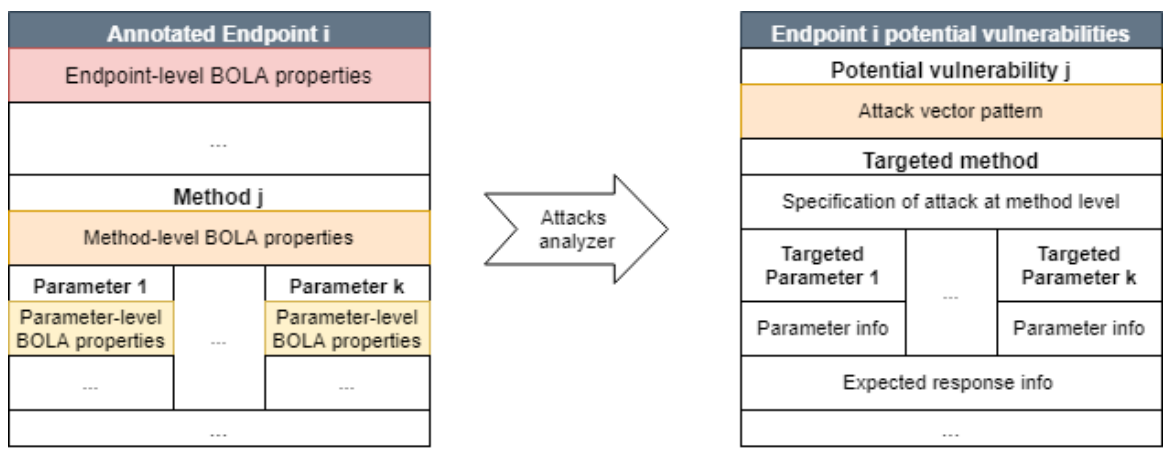

*Figure 4. Annotated endpoint description converts to a list of potential vulnerabilities for that endpoint with specified attack vectors proposals*

1) Annotated OpenAPI specification (stage "IDOR/BOLA properties analyzer" output) given as an input.
2) Determine order of individual endpoint processing based on corresponding path patterns, i.e. /{path}. Proceed with analysis of the first endpoint's path item object.
3) For each attack vector pattern:
    a. Read condition it requires for applicability. Let's consider it splits into conditions for endpoint level attributes, method level attributes and parameter level attributes.
    b. If endpoint level properties' values do not satisfy conditions then go to the next attack vector pattern.
    c. For each method specified for that endpoint:
        i. If corresponding method level properties' values do not satisfy conditions then go to the next endpoint's method.

ii. For each parameter used in the method, check if its parameter value properties satisfy the condition. If all conditions are satisfied, potential vulnerability is detected and the attack vector pattern is specified with corresponding endpoint, methods and parameters.

For some attack vector patterns process may be slightly modified in a way that iterating is done over a pair of methods (like in Verb Tampering attack) or pair of parameters (like in Prototype Pollution).

Because IDOR/BOLA-related attack vectors use resource ID manipulation we need to automatically identify parameters in OpenAPI specification that are used as resource ID. It should be noted that developers could use any name they want and there are no special parameters in OpenAPI specification to identify resource ID. Therefore, to identify where or not parameter is resource ID, we use the following heuristic rules (Table 4).

*Table 4*

| Rule group | Rule definition |
|---|---|
| Based on parameter's name | Ends with "ID", "_id" or "Id" (based on different naming conventions) |
| | Ends with "UUID" or "GUID" in any case |
| | Equals "ID", "UUID", "GUID" in any case |
| | Parameter's name is reasonably an identifier's name, e.g. "name", "filename", "group", "key", "phone", "email" etc. A dictionary of common identifier's name is constructed to look up on it. |
| Parameter is produced by some other endpoint of that service | It is common pattern to compose endpoint's path in a way that, for instance, /buckets path of an endpoint creates resource and produces resource's ID written in a response. In the same time /buckets/{bucketID} endpoint consumes bucketID identifier |
| Based on parameter location and endpoint's patterned path | Parameter's location is path and patterned path contains preceding string close to parameter's name. For example, /accounts/{account} - account is an identifier's name and preceding 'accounts' string contains 'account' substring. |
| | Parameter's location is path and patterned path contains two preceding string - the first is close to parameter's name and the second is verb representing an action. For instance, /profileInfo/edit/{profile} |
| | Parameter's location is path and at the start of patterned path string, e.g. /{collection}/action/... |
| Based on parameter's description | parameter's 'description' field is a string that contains words like "ID", "UUID", "GUID" or "identifier" (case insensitive) |
| Based on tags ussage | Usage of tags for parameter or parameter's schema: 'tags' field has a string that contains words like "ID", "UUID", "GUID" or "identifier" (case insensitive) |

In order to determine identifier's type, we use the following heuristic rules:

- Field 'type' of parameter's 'schema' is integer – then is type "integer";
- Field 'type' of parameter's 'schema' is string and identifier's name contains "UUID" or "GUID" – then is type "UUID/GUID";
- Field 'type' of parameter's 'schema' is array – then is type "array";
- Field 'type' of parameter's 'schema' is string and identifier's name contains words like 'email' 'phone' etc. – then is type "personal information";
- Field 'type' of parameter's 'schema' is string and none of the rules above is triggered - then is type "string";
- If none of the rules is triggered – then is type "other".

Let's consider the following example. As an input, we have the following OpenAPI specification (sample). Initial specification file, some parts are omitted for the sake of clarity (Figure 5).

```yaml
openapi: 3.0.2
paths:
  "/vaults/{vaultUuid}":
      get:
        operationId: GetVaultById
        parameters:
          - description: The UUID of the Vault to fetch Items from
            in: path
            name: vaultUuid
            required: true
            schema:
              pattern: ^[\da-z]{26}$
              type: string
        responses:
          "200":
            ...
            description: OK
          "401":
            ...
            description: Invalid or missing token
          "403":
            ...
            description: Unauthorized access
          "404":
            ...
            description: Vault not found
        security:
          - ConnectToken: []
        summary: Get Vault details and metadata
```

*Figure 5 Valid input OpenAPI specification in YAML format.*

IDOR/BOLA analyzer processes specification to annotate with properties, IDOR/BOLA endpoint analysis annotations output (Figure 6, cursive)

```yaml
openapi: 3.0.2
paths:
  /vaults/{vaultUuid}:
    get:
      operationId: GetVaultById
      parameters:
      - description: The UUID of the Vault to fetch Items from
        in: path
        name: vaultUuid
        required: true
        schema:
          pattern: ^[\da-z]{26}$
          type: string
        parameter_level_properties:
          is_identifier: true
          location: path
          type: UUID
      responses:
        ...
      security:
      - ConnectToken: []
      summary: Get Vault details and metadata
      tags:
      - Vaults
      method_level_properties:
        operation_only_parameters_specified: true
        parameters_required: true
```

```
      has_body: false
      identifiers_used: single
      authorization_required: true
  endpoint_level_properties:
    defined_http_verbs: Single
```

*Figure 6. Annotated OpenAPI specification*

Annotated OpenAPI specification is analyzed to outline potential attacks (Figure 7).

```
/vaults/{vaultUuid}:
  count: 3
  attacks:
  - …
  - …
  - name: Enumeration
    check_rule: Operation uses parameters AND operation has identifiers
    description: 'Identifier is tampered for enumeration based on automatically or
      semi-automatically determined pattern. In the simplest form, identifier is sequential
      and enumeration leads to targeting existing object with identifier being unknown
      at the start '
    targeted_operation: get
    targeted_parameters:
      vaultUuid:
        attacks:
        - Enumeration with a priori knowledge
        403_response_code_specified: true
        parameter_level_properties:
          is_identifier: true
          location: path
          type: UUID
        additional_check_rule: Identifier's type is UUID
    expected_response:
      content:
        application/json:
          example:
            message: vault {vaultUuid} is not in scope
            status: 403
          schema:
            $ref: '#/components/schemas/ErrorResponse'
      description: Unauthorized access
    unexpected_response_codes:
    - '200'
    - '401'
    - '404'
```

*Figure 7. Proposed attacks for the endpoint based on potential vulnerabilities found*

# 5 Experimental validation

Experimental validation of the developed approach should reflect applicability of the algorithmic advantage into real application security engineer workflow. During analysis of applicability it is required to demonstrate in experiments applicability for detection of potential access control vulnerabilities. Applicability testing requires experiments in two directions that describe real scenarios as coverage of vulnerabilities of all types in specifications and processing of specifications that contain vulnerabilities of different types.

In the first scenario, it is necessary to test identification and test generation for rather simple specifications that contain exact type and generated test is relevant to use to simplify security engineering and reduce time for generation of test this type scenarios. Such scenarios should be prepared for each vulnerability type and resulting test relevance should be verified as expected.

Second scenario should process close to real specification with a large number of endpoints. Solution should provide and verify expected number of endpoints and all detected potential vulnerabilities lead to generated relevant tests of corresponding type. Results should be verified using classic instruments set that are used by security engineers with clarification of final output statistics.

For the first experiment scenario during research we generated specification examples that contain at least one vulnerability of each type from the table 3. These example processing is expected to lead to potential access control vulnerability discovery by analyzing modules of proposed solution and corresponding test case generation. These simple and obvious examples verify that the solution covers test generation of all required types or some subset of known types in case some test examples result varies from expected.

Second step in this scenario is generation of specification that contains a predefined number of endpoints we are able to verify in total number of processed endpoints, between generated endpoints we generate endpoints with potential access control vulnerabilities of several types that can be verified in total types of detected vulnerabilities. As a final test output we verify the number of generated tests for the generated specification according to generation parametrization. Therefore, if we generate specifications with 2 endpoints containing vulnerability of type 1, 3 endpoints with vulnerability of type 2 and 5 endpoints with no vulnerabilities we expect in the output 10 processed endpoints, discovered 2 types of potential access control vulnerabilities and generated 5 tests for them. This would verify that the solution can process full specifications that contain several numbers of potential access control vulnerabilities and the result of output is verified by specification generation parameters.

Second step of the first scenario would guarantee that during real specification processing results would generate tests for all detected potential access control vulnerabilities discovered during processing of full specification.

For the second scenario, we use public available specifications that contain potential vulnerabilities. According to the test scenario during such specification examples processing should lead to verification of several criteria:

- All available endpoints were processed by the proposed solution and the number of processed endpoints is verified to be the same as expected. Expected numbers can be verified by manual tools for endpoints number verification.
- All discovered potential access control vulnerabilities listed in the output and corresponding tests can be verified as a vulnerability of appropriate type and verified by security engineer and generated tests verified as relevant for each vulnerability. This is a kind of true positive example verification.
- For a real specification where the exact number of potential access control vulnerabilities is undefined that's why can't be verified even by a human expert but it is possible to exclude from the real specification all endpoints that were signed as having vulnerability and tests were already generated. The rest of specification without already discovered endpoints can be verified by experts as an endpoint without vulnerability of this type. In case of too many endpoints left for manual analysis without any type of potential access control vulnerabilities discovered we can pick a random subset of such endpoints for manual verification. This would provide us with an assessment of percent of False Negative results in the output.

Our experimental results are summarized in the tables below (Table 5 – Table 8).

*Table 5*

| Test case ID | # Endpoints in specification | # Processed Endpoints | Endpoint processing ratio, % | # Methods in specification | # Processed Methods | Method processing ratio, % |
|---|---|---|---|---|---|---|
| | | | | | | |

| Test_1 | 11  | 11  | **100** | 15  | 15  | **100** |
| Test_2 | 74  | 74  | **100** | 111 | 111 | **100** |
| Test_3 | 25  | 25  | **100** | 48  | 48  | **100** |
| Test_4 | 22  | 22  | **100** | 31  | 31  | **100** |
| Test_5 | 25  | 25  | **100** | 25  | 25  | **100** |
| Test_6 | 182 | 182 | **100** | 304 | 304 | **100** |
| Test_7 | 6   | 6   | **100** | 8   | 8   | **100** |

*Table 6*

| Test case ID | # Identifiers in specification | #True Identifiers recognised | #False identifiers recognised | True positive, % | False Positive, % |
|---|---|---|---|---|---|
| Test_1 | 3  | 3  | 0 | **100** | 0 |
| Test_2 | 23 | 18 | 1 | **78,3** | 4,3 |
| Test_3 | 11 | 11 | 1 | **100** | 9,1 |
| Test_4 | 2  | 2  | 0 | **100** | 0 |
| Test_5 | 8  | 5  | 1 | **62,5** | 12,5 |
| Test_6 | 45 | 39 | 0 | **86,7** | 0 |
| Test_7 | 2  | 2  | 0 | **100** | 0 |

*Table 7*

| Test case ID | # Generated tests | # Relevant tests | True positive (TP), % |
|---|---|---|---|
| Test_1 | 22  | 22  | **100** |
| Test_2 | 194 | 192 | **98,9** |
| Test_3 | 73  | 64  | **87,6** |
| Test_4 | 84  | 76  | **90,5** |
| Test_5 | 65  | 64  | **98,5** |
| Test_6 | 564 | 564 | **100** |
| Test_7 | 16  | 12  | **75** |

*Table 8*

| Test case ID | # Endpoints with tests generated | # Endpoint without tests | # Endpoints without any generated tests but do have potential vulnerabilities | False Negative, % |
|---|---|---|---|---|
| Test_1 | 8   | 3  | 0 | **0** |
| Test_2 | 55  | 19 | 2 | **10,5** |
| Test_3 | 18  | 7  | 0 | **0** |
| Test_4 | 22  | 0  | 0 | **0** |
| Test_5 | 25  | 0  | 0 | **0** |
| Test_6 | 178 | 4  | 0 | **0** |
| Test_7 | 6   | 0  | 0 | **0** |

Our paper has made the first step towards automatic detection of potential IDOR/BOLA-related vulnerabilities based on API specification processing during threat modeling (architecture design) phase.

However, it is still preliminary and has several limitations for future improvement, which we discuss below.

- **Availability of the API specification**. In this paper, we assume that security assessment team can get API specification written in OpenAPI format. However, there may be a case where OpenAPI specification is not available – in that case security assessment team needs to review application source code and extracts information about endpoints such path, consumed and produced parameters.
- **Trustworthiness of the API specification**. If API specification is available we assume that the specification is benign. E.g., development team define "securitySchemes" object and properly annotate each operation with the required security schemes in case operation requires any security key to process request.
- **Our tool output is potential vulnerabilities, not confirmed.** Proposed method and tool are not provide any dynamic security testing, so the output is potential vulnerabilities, not confirmed.

# 6. Related work

**API specification usage for security purposes.** Atlidakis at al. [1], [2] proposed stateful REST API fuzzing technique and tool RESTler. RESTler analyzes the REST API specification (in OpenAPI format) of a service under the test, generates operation dependency graphs (i.e., producer-consumer dependencies among requests) and then generates sequences of requests that automatically test the service through its API. Similarly to Atlidakis at al. approach, Viglianisi at al. [3] model the dependencies among the operations in a REST API to elaborate an appropriate ordering. In contrast with Atlidakis at al. approach, they propose to compute the next operations to test dynamically, based on the outcome of the operations that could be tested so far. S. Karlsson at al. [4], [5] proposed a method and tool that for a given OpenAPI specification, produces input generators that are used in property-based tests as well as produces automatic oracles to find software bugs during dynamic testing. Laranjeiro at al. [6] proposed to use OpenAPI specification for carrying out robustness tests on REST APIs in order to identify and extract relevant information (e.g., unique API endpoint URIs) for testing the service without operation dependency graph generation. Then such information is used to generate payload. Arcuri at al. [7] used OpenAPI specification to automatically generate valid HTTP calls and provide automatic dynamic testing of REST API in black-box and white-box modes. 42Crunch API Security Audit tool takes OpenAPI specification as an input, provides its static analysis and generates the set of security alters (without focus on potential IDOR/BOLA) based on the set of predefined rules. In summary, currently OpenAPI specification is used to generate valid HTTP requests or sequences of HTTP requests during dynamic software testing in order to find bugs and vulnerabilities or in static mode to find security-specific issues with specification without focus on IDOR/BOLA. In this paper we propose to provide static analysis of OpenAPI specification (without dynamic software testing) to identify potential IDOR/BOLA vulnerabilities and automatically derive non-functional security requirements to mitigate those threats. To use the proposed method we need to obtain OpenAPI specification, without access to the system under the test source code, binary files and testing environment.

**Application security threat modeling.** One of the common approach to provide application security threat modeling during application development is to use data flow diagrams [9], [10], [11], [14] or other formal notations [20] to describe an application under the threat modeling and then derive a set of threats and corresponding mitigations techniques. Another direction is to derive non-functional security requirements based on existing sets of attack patterns like MITRE ATT&CK and CAPEC [12], [13]. There are also several tools aimed to automate that process, e.g. OWASP Threat Dragon, threatspec. In contrast with existing works and tools, our approach does not need any additional artifacts (like data flow diagrams) as input and generates precise non-functional requirements ready to include in the development process.

**IDOR/BOLA vulnerability detection methods and tools.** There are limited numbers of research works in academia aimed to design an approach to detect access control vulnerabilities, especially IDOR/BOLA. Atlidakis at al. [2] proposed an IDOR/BOLA checker based on stateful REST API fuzzing technique and OpenAPI specification parsing to find IDOR/BOLA via dynamic testing. Viriya at al. [8] proposes an IDOR/BOLA testing methodology that is not based on information extraction from API specification. There are also several tools (RESTler from Microsoft research and fuzz-lightyear from Yelp) aimed to automate the IDOR/BOLA detection process in dynamic testing. Those tools take OpenAPI specification as input and apply stateful REST API fuzzing techniques to detect vulnerabilities. In contrast with existing works and tools, our approach is aimed to use at the threat model phase and need just OpenAPI specification as input. In the other case, because our approach does not do any dynamic testing, the out is the list of potential IDOR/BOLA vulnerabilities, not confirmed.

# 7. Conclusion

We have presented the first algorithm for automatic detection of IDOR/BOLA-related vulnerabilities in application based on API specification. It includes two fundamental techniques: (1) an OpenAPI static analysis and annotation its properties related with potential IDOR/BOLA-related vulnerabilities, and (2) a rule-based attack analyzer that process annotated OpenAPI specification and produces potential attack vectors description. We have implemented a prototype and tested it with seven popular microservice applications. Our experimental results show that proposed algorithm can effectively detect potential IDOR/BOLA-related vulnerabilities. Proposed algorithm can be used in different SDL phases [21]: (1) in design phase to design and provide development team during sprint planning with concreate set of non-functional requirements aimed to mitigate IDOR/BOLA-related vulnerabilities; (2) in testing phase to design the set of penetration tests to detect IDOR/BOLA-related vulnerabilities; (3) in operational phase to design attack detection techniques.

# Acknowledgments

We thank Anatoly Katyushin and Denis Valeev for the assistance and many constructive discussions.

# АВТОМАТИЧЕСКОЕ ВЫЯВЛЕНИЕ УЯЗВИМОСТЕЙ, СВЯЗАННЫХ С НЕДОСТАТКАМИ УПРАВЛЕНИЯ ДОСТУПОМ, НА ОСНОВЕ АНАЛИЗА СПЕЦИФИКАЦИИ ВНЕШНИХ ПРОГРАММНЫХ ИНТЕРФЕЙСОВ

Барабанов А.В.[10], Дергунов Д.О.[11], Макрушин Д.Н.[12], Теплов А.М.[13]

**Аннотация**

**Цель статьи**.

Уязвимости, связанные с недостатками управления доступом и защиты данных (Insecure Direct Object Reference/IDOR или Broken Object Level Authorization/BOLA), являются одним из критических типов уязвимостей современных веб-приложений, использование которых может привести к нарушениям конфиденциальности и целостности защищаемой информации. Цель проведенного исследования заключалась в разработке алгоритма, позволяющего повысить эффективность выявления подобных уязвимостей при разработке безопасного программного обеспечения и сократить время проведения тестирования на проникновения.

**Метод исследования** заключается в системном анализе научных публикаций, выступлений на ведущих научно-технических конференциях, отчетов программ вознаграждения за найденные уязвимости и других неофициальных источников, обобщении полученных результатов с целью систематизации сведений о компьютерных атаках, направленных на эксплуатацию уязвимостей, связанных с недостатками управления доступом. При разработке алгоритма использовался метод статического анализа, для оценки эффективности разработанного алгоритма проводилось функциональное тестирование.

**Полученные результаты и практическая значимость**. Представлена систематизация сведений о компьютерных атаках, направленных на эксплуатацию уязвимостей, связанных с недостатками управления доступом и защиты данных. Предложен алгоритм автоматического формирования описания тестов выявления подобных уязвимостей на основе статического анализа спецификации в формате OpenAPI внешних интерфейсов веб-приложения и реализация алгоритма на языке программирования Python. Проведенная оценка эффективности показала, что разработанный алгоритм может быть использован на практике при проведении тестирований на проникновение на этапе разработки описания тестов.

## Литература

---

[10] Александр Владимирович Барабанов, канд.техн.наук, CISSP, CSSLP, доцент кафедры «Информационная безопасность» (ИУ-8) МГТУ им. Н.Э. Баумана. E-mail: barabanov.iu8@gmail.com

[11] Денис Олегович Дергунов, инженер, Лаборатория передовых программных технологий, компания Huawei. E-mail: dan.dergunow@gmail.com

[12] Денис Николаевич Макрушин, OSCP, директор, Центр программного обеспечения и технологий тестирования Chong-Ming, компания Huawei. E-mail: denis@makrushin.com

[13] Алексей Михайлович Теплов, канд.техн.наук, главный инженер программного обеспечения, Лаборатория передовых программных технологий, компания Huawei. E-mail: aleksey-teplov@huawei.com

[6] N. Laranjeiro, J. Agnelo and J. Bernardino, "A Black Box Tool for Robustness Testing of REST Services," in IEEE Access, vol. 9, pp. 24738-24754, 2021, doi: 10.1109/ACCESS.2021.3056505.

[7] A. Arcuri, "Automated Black- and White-Box Testing of RESTful APIs With EvoMaster," in IEEE Software, vol. 38, no. 3, pp. 72-78, May-June 2021, doi: 10.1109/MS.2020.3013820.

[8] Viriya, Anthony, and Yohan Muliono. "Peeking and Testing Broken Object Level Authorization Vulnerability onto E-Commerce and E-Banking Mobile Applications." Procedia Computer Science 179 (2021): 962-965.

[9] Laurens Sion, Koen Yskout, Dimitri Van Landuyt, and Wouter Joosen. 2018. Solution-aware data flow diagrams for security threat modeling. In Proceedings of the 33rd Annual ACM Symposium on Applied Computing (SAC '18). Association for Computing Machinery, New York, NY, USA, 1425–1432. DOI:https://doi.org/10.1145/3167132.3167285

[10] Laurens Sion, Koen Yskout, Dimitri Van Landuyt, Alexander van den Berghe, and Wouter Joosen. 2020. Security Threat Modeling: Are Data Flow Diagrams Enough? In Proceedings of the IEEE/ACM 42nd International Conference on Software Engineering Workshops (ICSEW'20). Association for Computing Machinery, New York, NY, USA, 254–257. DOI:https://doi.org/10.1145/3387940.3392221

[11] Katja Tuma, Laurens Sion, Riccardo Scandariato, and Koen Yskout. 2020. Automating the early detection of security design flaws. In Proceedings of the 23rd ACM/IEEE International Conference on Model Driven Engineering Languages and Systems (MODELS '20). Association for Computing Machinery, New York, NY, USA, 332–342. DOI:https://doi.org/10.1145/3365438.3410954

[12] Seungoh Choi, Jeong-Han Yun, and Byung-Gil Min. 2021. Probabilistic Attack Sequence Generation and Execution Based on MITRE ATT&CK for ICS Datasets. In Cyber Security Experimentation and Test Workshop (CSET '21). Association for Computing Machinery, New York, NY, USA, 41–48. DOI:https://doi.org/10.1145/3474718.3474722

[13] M. Vanamala, J. Gilmore, X. Yuan and K. Roy, "Recommending Attack Patterns for Software Requirements Document," 2020 International Conference on Computational Science and Computational Intelligence (CSCI), 2020, pp. 1813-1818, doi: 10.1109/CSCI51800.2020.00334.

[14] C. Wilhjelm and A. A. Younis, "A Threat Analysis Methodology for Security Requirements Elicitation in Machine Learning Based Systems," 2020 IEEE 20th International Conference on Software Quality, Reliability and Security Companion (QRS-C), 2020, pp. 426-433, doi: 10.1109/QRS-C51114.2020.00078.

[15] Fredj O.B., Cheikhrouhou O., Krichen M., Hamam H., Derhab A. (2021) An OWASP Top Ten Driven Survey on Web Application Protection Methods. In: Garcia-Alfaro J., Leneutre J., Cuppens N., Yaich R. (eds) Risks and Security of Internet and Systems. CRiSIS 2020. Lecture Notes in Computer Science, vol 12528. Springer, Cham. https://doi.org/10.1007/978-3-030-68887-5_14

[16] Pereira-Vale, A., Fernandez, E.B., Monge, R., Astudillo, H., Márquez, G., 2021. Security in microservice-based systems: A multivocal literature review. Comput. Secur. 103, 102200. https://doi.org/10.1016/j.cose.2021.102200

[17] Chris Richardson, "Microservices patterns: with examples in Java," ed. Switzerland, Europe: Manning Publications, 2019.

[18] F. Osses, G. Marquez, and H. Astudillo, "An exploratory study of academic architectural tactics and patterns in microservices: A systematic literature review," in Avances en Ingenieria de Software a Nivel Iberoamericano, CIbSE 2018, 2018.